\documentclass[a4paper, 10pt]{article}

\usepackage[latin1]{inputenc}
\usepackage{subfigure}
\usepackage{amsmath}
\usepackage{amssymb}
\usepackage{amsthm}
\usepackage{amsfonts}
\usepackage[cyr]{aeguill}
\usepackage{xspace}
\usepackage[english]{babel}
\usepackage{graphicx}
\usepackage{float}
\usepackage{mathtools}
\usepackage{psfrag,floatflt,hyperref}
\usepackage{verbatim}
\usepackage{geometry}
\usepackage{multirow}
\usepackage[retainorgcmds]{IEEEtrantools}
\usepackage{stmaryrd}
\SetSymbolFont{stmry}{bold}{U}{stmry}{m}{n}
\usepackage{arydshln,leftidx,mathtools}
\usepackage{color}

\newtheorem*{theorem*}{Theorem}

\title{Low temperature ratchet current\footnote{The author acknowledges support from the Swiss NSF grant $200021\_132528/1$.}}
\author{Justine Louis}
\date{$5$th August $2015$}

\newcommand*{\defeq}{\mathrel{\vcenter{\baselineskip0.5ex \lineskiplimit0pt
                     \hbox{\scriptsize.}\hbox{\scriptsize.}}}=}

\begin{document}
        \maketitle

\begin{abstract}
In \cite{maes2014low}, the low temperature ratchet current in a multilevel system is considered. In this note, we give an explicit expression for it and find its numerical value as the number of states goes to infinity.
\end{abstract}

\section{Introduction}
In this note, we compute the stationary ratchet current in the large system size limit. In \cite{maes2014low}, the authors derive a formula for the occupation of a general multilevel system at low temperature. As an application, they consider a continuous time version of Parrondo's game at low temperature (see \cite{parrondo1998reversible}) and give an expression for the ratchet current. We consider a multilevel system determined by a finite number of states. The set of all states is denoted by $K$. The ratchet is modelised by two rings of $N$ states. In the present section, we recall the definitions and results from \cite[section $3$]{maes2014low} and in the next section we give an explicit expression for the ratchet current using the Tutte matrix tree theorem and find its limit as the number of states goes to infinity. The states on the outer ring are denoted by $(0,i)$ and on the inner ring by $(1,i)$, where $i=1,\ldots,N$. The energies are denoted by $E_i$, $i=1,\ldots,N$ and are such that $E_1<\cdots<E_N$. The transition rates on the outer ring are given by
\begin{equation*}
\lambda((i,0),(i+1,0))=e^{\beta(E_i-E_{i+1})/2},\quad\lambda((i+1,0),(i,0))=e^{\beta(E_{i+1}-E_i)/2}
\end{equation*}
where $\beta$ is the inverse temperature. On the inner ring, the transition rates are constant and equal to one, that is,
\begin{equation*}
\lambda((i,1),(i+1,1))=\lambda((i+1,1),(i,1))=1.
\end{equation*}
The two rings are connected with transition rates constant equal to one,
\begin{equation*}
\lambda((i,n),(i,1-n))=1,\textnormal{ where }n=0,1.
\end{equation*}
The zero-temperature logarithmic limit denoted by $\phi(x,y)$ is given by
\begin{equation*}
\phi(x,y)\defeq\lim_{\beta\rightarrow\infty}\frac{1}{\beta}\log{\lambda(x,y)}.
\end{equation*}
The zero-temperature logarithmic limit of the escape rates of state $x$ is denoted by $\Gamma(x)$ and given by
\begin{equation*}
\Gamma(x)\defeq-\lim_{\beta\rightarrow\infty}\frac{1}{\beta}\log\big(\sum_y\lambda(x,y)\big)=-\max_y\phi(x,y).
\end{equation*}
The logarithmic-asymptotic transition probability is given by $e^{-\beta U(x,y)}$ where
\begin{equation*}
U(x,y)\defeq-\Gamma(x)-\phi(x,y).
\end{equation*}
We have $U(x,y)\geqslant0$ for all $x,y\in K$. The smaller $U(x,y)$ is, the larger is the probability of transition from state $x$ to state $y$. Hence, the set of preferred successors of $x$ is defined by
\begin{equation*}
\{y\in K\mid U(x,y)=0\}.
\end{equation*}
When $U(x,y)=0$, the probability of transition from $x$ to $y$ is high. Thus we consider the directed graph $K^D$ defined by the vertex set $K$ and edge set $\{(x,y)\mid U(x,y)=0\}$ where $(x,y)$ indicates an oriented edge from $x$ to $y$. The digraph $K^D$ is represented in Figure \ref{rc} below. The low temperature asymptotic of the stationary occupation is given in the following theorem from \cite{maes2014low}:
\begin{figure}[!ht]
\centering
\includegraphics[scale=0.55]{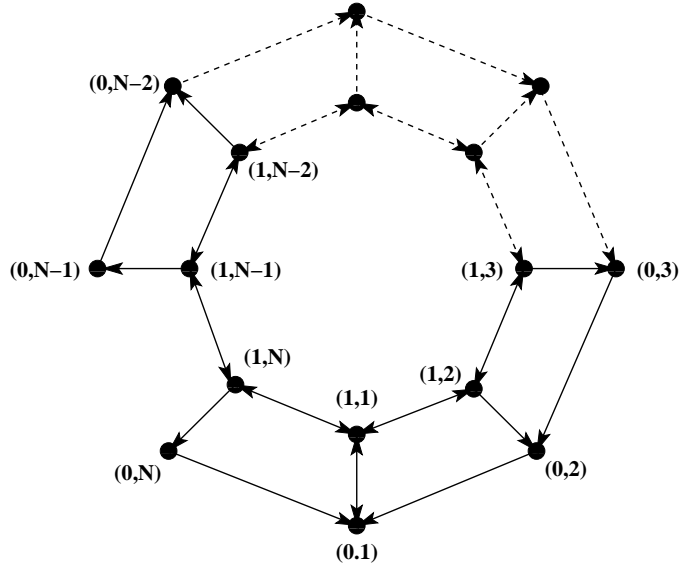}
\caption{The directed graph $K^D$.}
\label{rc}
\end{figure}
\begin{theorem*}[\protect{\cite[Theorem $2.1$]{maes2014low}}]
There is $\epsilon>0$ so that as $\beta\rightarrow\infty$,
\begin{equation*}
\rho(x)=\frac{1}{\mathcal{Z}}A(x)e^{\beta(\Gamma(x)-\Theta(x))}(1+O(e^{-\beta\epsilon}))
\end{equation*}
with
\begin{align*}
&\Theta(x)\defeq\min_{\mathcal{T}}U(\mathcal{T}_x)\quad\textnormal{for}\quad U(\mathcal{T}_x)\defeq\sum_{(y,y')\in\mathcal{\mathcal{T}_x}}U(y,y')\quad\textnormal{and}\\
&A(x)\defeq\sum_{\mathcal{T}\in M(x)}\prod_{(y,y')\in\mathcal{T}_x}a(y,y')=e^{o(\beta)}
\end{align*}
where the last sum runs over all spanning trees minimizing $U(\mathcal{T}_x)$ (i.e. $\mathcal{T}\in M(x)$ if $\Theta(x)=U(\mathcal{T}_X)$), and $a(x,y)$ are the reactivities, which are the sub-exponential part of the transition rates $\lambda(x,y)$.
\end{theorem*}
Here all the reactivities are constant equal to one, $a(x,y)=1$ for all $x,y\in K$. In the present case, for all $x\in K$, there exists an in-spanning tree $\mathcal{T}_x$ in $K^D$, so that $U(\mathcal{T}_x)=0$, and therefore $\Theta(x)=0$. Let $\mathcal{D}$ be the set of states for which $\Gamma(x)=0$, it is given by $\mathcal{D}=\{(1,0),(i,1),i=1,\ldots,N\}$. We denote $f\simeq g$ if $f=g+O(e^{-\beta\epsilon})$ as $\beta\rightarrow\infty$. For $x\in\mathcal{D}$, we have $\rho(x)\simeq\lvert M(x)\rvert/\mathcal{Z}$, where $\lvert M(x)\rvert$ is the number of in-spanning trees in $K^D$. For $x\notin\mathcal{D}$, the stationary distribution is exponentially small since from the theorem it is given by $\rho(x)\simeq\lvert M(x)\rvert e^{\beta\Gamma(x)}/\mathcal{Z}$, with $\Gamma(x)<0$. The stationary ratchet current in the clockwise direction is given by
\begin{equation*}
J_R=j((i+1,0),(i,0))+j((i+1,1),(i,1)),\quad\textnormal{for}\quad i=1,\ldots,N,
\end{equation*}
where
$j(x,y)=\lambda(x,y)\rho(x)-\lambda(y,x)\rho(y)$.\\
For $i=1$,
\begin{equation*}
J_R=j((2,0),(1,0))+j((2,1),(1,1)).
\end{equation*}
On the outer ring, we have $j((2,0),(1,0))=\lambda((2,0),(1,0))\rho(2,0)-\lambda((1,0),(2,0))\rho(1,0)$ with
\begin{align*}
&\lambda((1,0),(2,0))\simeq0,\quad\lambda((2,0),(1,0))=e^{(E_2-E_1)\beta/2}\\
&\rho(2,0)\simeq\frac{\lvert M(2,0)\rvert}{\mathcal{Z}}e^{\beta\Gamma(2,0)}=\frac{\lvert M(2,0)\rvert}{\mathcal{Z}}e^{-(E_2-E_1)\beta/2},
\end{align*}
so that $j((2,0),(1,0))\simeq\lvert M(2,0)\rvert/\mathcal{Z}$.\\
On the inner ring, we have $j((2,1),(1,1))=\lambda((2,1),(1,1))\rho(2,1)-\lambda((1,1),(2,1))\rho(1,1)$ with
\begin{align*}
&\lambda((2,1),(1,1))=\lambda((1,1),(2,1))=1,\\
&\rho(2,1)\simeq\frac{\lvert M(2,1)\rvert}{\mathcal{Z}},\quad\rho(1,1)\simeq\frac{\lvert M(1,1)\rvert}{\mathcal{Z}},
\end{align*}
so that $j((2,1),(1,1))\simeq(\lvert M(2,1)\rvert-\lvert M(1,1)\rvert)/\mathcal{Z}$. The ratchet current is thus given by
\begin{equation*}
J_R\simeq\frac{1}{\mathcal{Z}}(\lvert M(2,0)\rvert+\lvert M(2,1)\rvert-\lvert M(1,1)\rvert).
\end{equation*}
Considering converging arborescences, the Laplacian matrix of a directed graph is defined by $L=D-A$ where $D$ is the diagonal out-degree matrix and $A=(A_{ij})$ is the adjacency matrix such that $A_{ij}$ is the number of directed edges from $i$ to $j$. The rows and columns of $L$ are indexed by the vertices of the graph. Here, we index it first by the states on the outer ring then the ones on the inner ring, that is $(1,0),(2,0),\ldots,(N,0),(1,1),(2,1),\ldots,(N,1)$. The Tutte matrix tree theorem (see \cite{MR2339282}) relates the number of spanning arborescences converging to $x$ in $K^D$ to the cofactors of the Laplacian $\det L_{x,y}$. Let $x\in K$. Then for all $y\in K$,
\begin{equation*}
\lvert M(x)\rvert=(-1)^{x+y}\det L_{x,y}.
\end{equation*}
In particular, for $y=x$, we have $\lvert M(x)\rvert=\det L_x$. Therefore we have
\begin{equation*}
J_R\simeq\frac{1}{\mathcal{Z}}(\det L_{(2,1)}+\det L_{(2,0)}-\det L_{(1,1)}).
\end{equation*}
The Laplacian matrix is given by
\begin{equation*}
L=\left(
    \begin{array}{c;{2pt/2pt}c}
        A&B\\ \hdashline[2pt/2pt]
        Id&C 
    \end{array}
\right)
\end{equation*}
where $A$ is the $N\times N$ lower triangular matrix given by
\begin{equation*}
A=\begin{pmatrix}
1&&&&\\
-1&1&&&\\
&\ddots&\ddots&&\\
&&-1&1&\\
-1&&&0&1
\end{pmatrix},
\end{equation*}
$B$ is the $N\times N$ matrix such that all coefficients are zero except $B_{(1,0),(1,1)}=-1$, the matrix $Id$ is the $N\times N$ identity matrix and $C$ is the following circulant matrix
\begin{equation*}
C=\begin{pmatrix}
3&-1&&&-1\\
-1&3&\ddots&&\\
&\ddots&\ddots&\ddots&\\
&&\ddots&\ddots&-1\\
-1&&&-1&3
\end{pmatrix}.
\end{equation*}
\section{Calculation of the ratchet current}
From \cite{maes2014low}, the numerator of $J_R$ is given by
\begin{equation*}
\det L_{(2,1)}+\det L_{(2,0)}-\det L_{(1,1)}=\det B_{N-1}-2\det B_{N-2}-2
\end{equation*}
where $B_N$ is the $N\times N$ tridiagonal matrix with $3$ on the diagonal and $-1$ on the two off-diagonals which satisfies the recurrence relation $\det B_N=3\det B_{N-1}-\det B_{N-2}$ with $\det B_1=3$ and $\det B_2=8$. By solving the associated characteristic equation, it comes
\begin{equation*}
\det B_N=\frac{5-3\sqrt{5}}{10}\bigg(\frac{3-\sqrt{5}}{2}\bigg)^N+\frac{5+3\sqrt{5}}{10}\bigg(\frac{3+\sqrt{5}}{2}\bigg)^N.
\end{equation*}
The normalisation factor is given by
\begin{equation*}
\mathcal{Z}=\sum_{x\in K}\sum_{\mathcal{T}_x}\prod_{(y,z)\in\mathcal{T}_x}\lambda(y,z)\simeq\sum_{x\in\mathcal{D}}\lvert M(x)\rvert=\sum_{x\in\mathcal{D}}\det L_x.
\end{equation*}
The sum is over the states in $\mathcal{D}$ since the contribution of the states which are not in $\mathcal{D}$ is exponentially damped. Therefore we have
\begin{equation}
\label{Z}
\mathcal{Z}\simeq\det L_{(1,0)}+\sum_{i=1}^N\det L_{(i,1)}.
\end{equation}
We have
\begin{equation*}
\det L_{(1,0)}=\det C.
\end{equation*}
The circulant matrix $C$ has eigenvalues given by $\mu_j=3-2\cos(2\pi j/N)$, $j=0,1,\ldots,N-1$ (see \cite{MR1271140}). Hence
\begin{equation*}
\det L_{(1,0)}=\prod_{j=0}^{N-1}(3-2\cos(2\pi j/N))=U_{N-1}^2(\sqrt{5}/2)
\end{equation*}
where $U_N$ is the Chebyshev polynomial of the second kind. Thus
\begin{equation}
\label{L(1,0)}
\det L_{(1,0)}=\bigg(\frac{3+\sqrt{5}}{2}\bigg)^N+\bigg(\frac{3-\sqrt{5}}{2}\bigg)^N-2.
\end{equation}
From the Tutte matrix tree theorem, the cofactor $(-1)^{N+i}\det L_{(i,1)}$ is equal to the number of converging arborescences to $(i,1)$ and is equal to the cofactor of the Laplacian where row $(i,1)$ and any column is removed. Since the only non-zero element of $B$ is in column indexed by $(1,1)$, we choose to remove that one, so that
\begin{equation}
\label{M(i,1)}
\lvert M(i,1)\rvert=(-1)^{(N+i)+(N+1)}\det L_{(i,1),(1,1)}=(-1)^{i+1}\det C_{(i,1),(1,1)}
\end{equation}
since $A$ is lower triangular.
On the other hand, by adding to the first column of $C$ all the other ones, we have
\begin{equation}
\label{det(C)}
\det C=\begin{vmatrix}
1&-1&&&-1\\
1&3&\ddots&&\\
&-1&\ddots&\ddots&\\
&&\ddots&\ddots&-1\\
1&&&-1&3
\end{vmatrix}=\sum_{i=1}^N(-1)^{i+1}\det C_{(i,1),(1,1)}.
\end{equation}
Putting equations (\ref{Z}), (\ref{L(1,0)}), (\ref{M(i,1)}) and (\ref{det(C)}) together, we have
\begin{equation*}
\mathcal{Z}\simeq2\det C=2\bigg(\frac{3+\sqrt{5}}{2}\bigg)^N+2\bigg(\frac{3-\sqrt{5}}{2}\bigg)^N-4.
\end{equation*}
Up to exponentially small corrections $e^{-\beta\epsilon}$, the ratchet current is given for all $N$ by
\begin{align*}
J_R\simeq&\left(\frac{5+3\sqrt{5}}{10}\bigg(\frac{3+\sqrt{5}}{2}\bigg)^{N-1}+\frac{5-3\sqrt{5}}{10}\bigg(\frac{3-\sqrt{5}}{2}\bigg)^{N-1}-\frac{5+3\sqrt{5}}{5}\bigg(\frac{3+\sqrt{5}}{2}\bigg)^{N-2}\right.\\
&\left.-\frac{5-3\sqrt{5}}{5}\bigg(\frac{3-\sqrt{5}}{2}\bigg)^{N-2}-2\right)\bigg/\bigg(2((3+\sqrt{5})/2)^N+2((3-\sqrt{5})/2)^N-4\bigg).
\end{align*}
As a consequence, in the large system size limit the current saturates and has the following limit
\begin{equation*}
\lim_{N\rightarrow\infty}J_R\simeq\frac{1}{2}-\frac{1}{\sqrt{5}}.
\end{equation*}
\par\vspace{\baselineskip}
\noindent
\textbf{Acknowledgements:} The author thanks Anders Karlsson for suggesting this problem to her.

\nocite{*}
\bibliographystyle{plain}
\bibliography{bibliography}

\end{document}